\documentclass[useAMS,usenatbib]{mnras}
\usepackage{longtable}
\usepackage{graphicx}
\usepackage{lscape}
\usepackage{rotating}

\newcommand{\Msun}{\mbox{$\mathrm{M}_{\odot}$}}

\newcommand{\Ion}[2]{#1{\,\scriptsize #2}}

\title[The age-metallicity relation]  {The age-metallicity relation in
  the  solar neighbourhood  from a  pilot sample  of white  dwarf-main
  sequence  binaries}

\author[A.  Rebassa-Mansergas et  al.]{A.  Rebassa-Mansergas$^{1}$, B.
  Anguiano$^{2}$, E.  Garc\'ia-Berro$^{1,3}$,  K.\,C.  Freeman$^{4}$, 
  \newauthor R. Cojocaru$^{1}$,    C.\,J.  Manser$^{5}$,  A.\,F.   Pala$^{5}$,
  B.\,T. G\"ansicke$^{5}$, X.\,-W. Liu$^{6,7}$\\
$^{1}$ Departament de F\'\i sica, Universitat Polit\`ecnica de
  Catalunya, c/Esteve Terrades 5, 08860 Castelldefels, Spain\\
$^{2}$  Department of  Physics  and  Astronomy, Macquarie  University,
  North Ryde, NSW 2109, Australia\\
$^{3}$ Institute for Space Studies of Catalonia, c/Gran Capit\`a 2--4,
  Edif. Nexus 201, 08034 Barcelona, Spain\\
$^{4}$ Research School of Astronomy and Astrophysics, Australian
  National University, Canberra, ACT 2611, Australia\\
$^{5}$ Department of Physics, University of Warwick, Coventry CV4 7AL,
 UK \\
$^{6}$ Kavli Institute for Astronomy and Astrophysics, Peking
 University, Beijing 100871, P.\,R.\,China\\
$^{7}$ Department of Astronomy, Peking University, Beijing 100871,
 P.\,R.\,China\\
}

\begin{document}
\date{Accepted 2016. Received 2016; in original form 2016}
\pagerange{\pageref{firstpage}--\pageref{lastpage}} \pubyear{2016}
\maketitle

\begin{abstract}
The  age-metallicity relation  (AMR)  is  a fundamental  observational
constraint for understanding how the  Galactic disc formed and evolved
chemically in  time.  However, there  is not  yet an agreement  on the
observational  properties  of  the  AMR for  the  solar  neighborhood,
primarily due to the difficulty in obtaining accurate stellar ages for
individual field stars. We have  started an observational campaign for
providing  the much  needed observational  input by  using wide  white
dwarf-main  sequence (WDMS)  binaries.  White  dwarfs are  ``natural''
clocks and can be used to  derive accurate ages.  Metallicities can be
obtained from the main sequence  companions.  Since the progenitors of
white dwarfs and  the main sequence stars were born  at the same time,
WDMS  binaries   provide  a  unique  opportunity   to  observationally
constrain in a robust  way the properties of the AMR.  In this work we
present  the AMR  derived from  analysing a  pilot sample  of 23  WDMS
binaries  and provide  clear observational  evidence for  the lack  of
correlation between age and metallicity at young and intermediate ages
(0--7 Gyrs).
\end{abstract}


\label{firstpage}

\section{Introduction}
\label{s-intro}

The observed  evolution of  stellar abundances as  a function  of age,
i.e. the age-metallicity  relation (AMR), is the fossil  record of the
chemical evolution  and enrichment history  of the Galactic  disc. The
AMR is hence a critical observational constraint to understand how the
Galactic  disc  formed  and  evolved chemically.   The  AMR  has  been
extensively studied in the past decades \citep[see the reviews of e.g.
][and references  therein]{freeman+blandhawthorn02-1, nomotoetal13-1}.
Whilst early observational studies found a correlation between stellar
ages      and     metallicity      in      the     solar      vicinity
\citep[e.g.][]{rocha-pintoetal00-1,  soubiranetal08-1},   more  recent
works show  a substantial scatter  in the relation, suggesting  that a
clear correlation between  the age and the metallicity  does not exist
\citep[e.g.][]{haywoodetal13-1,       bergemannetal14-1}.        These
discrepancies in the derived AMRs are  likely to arise due to the fact
that measuring precise stellar ages  is a difficult endeavour, subject
to substantial uncertainties \citep{soderblom10-1}.  Indeed, important
differences between different works  arise when comparing ages derived
for  the same  stars \citep{anguianoetal10-1}.   The discrepancies  in
stellar ages, and  hence in the derived AMRs, are  the main motivation
to explore  other dating methods  and their application  to understand
the chemical evolution of the Galactic disc.

\begin{figure*}
\centering
\includegraphics[angle=-90,width=\textwidth]{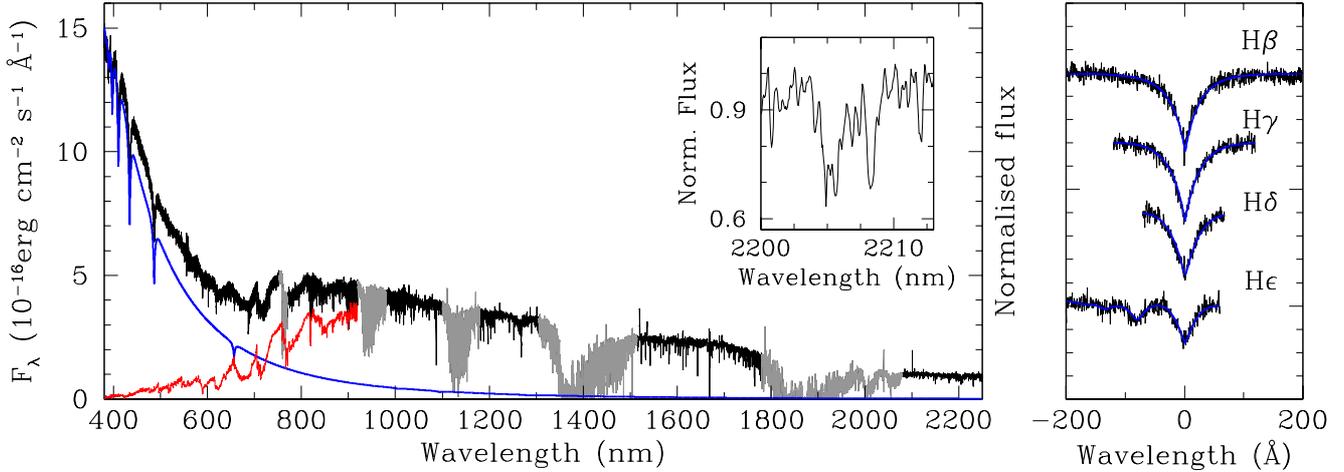}
\caption{Left panel:  example X-Shooter spectrum  of SDSSJ\,0036+0700,
  an  SDSS WDMS  binary in  our  sample (black;  regions dominated  by
  telluric absorption are  shown in gray).  The WD  dominates the flux
  contribution  at   blue  wavelengths  ($\lambda<600$\,nm),   the  MS
  companion contributes  most to the  red (and near-infrared)  part of
  the spectrum.   The best-fit WD  model and  MS star template  to the
  observed optical spectrum are shown as blue and red solid lines.  On
  the  top  right corner  we  zoom-in  to the  \Ion{Na}{I}  absorption
  doublet  at 2205/2209  nm  (free of  telluric  absorption), used  to
  derive the  MS star [Fe/H]  abundance.  Right panel:  the normalised
  residual WD Balmer  lines (black) that result  from substracting the
  MS  star contribution  and the  best-fit WD  model (blue).   The fit
  provides very  accurate values of  the WD effective  temperature and
  surface gravity, which are used to calculate the WD masses and total
  ages of the binaries.}
  \label{f-spectrum}
\end{figure*}

Open  clusters have  been frequently  used  to study  the AMR.   These
studies show  the same scatter  of metallicities as observed  in field
stars  \citep{friel95-1,  carraroetal98-1,  pancinoetal10-1}.   Hence,
these results  support the idea of  a lack of correlation  between the
metallicity and  the age.  However,  it is  important to bear  in mind
that    open     clusters    generally    dissolve     after    1\,Gyr
\citep{vandeputteetal10-1}.  This makes  it difficult  to address  the
study of the  AMR for intermediate and old ages.   Moreover, as argued
by \citet{casamiquelaetal16-1},  the lack  of a  large number  of open
clusters analysed  homogeneously hampers the investigations  about the
existence of an  AMR, which may leave these  results inconclusive.  In
this  paper we  aim  at providing  additional  observational input  to
address  the  existence   of  the  AMR  using  a   different  kind  of
astronomical ``clock'': white dwarfs.

White dwarfs (WDs)  are the most typical endpoint of  the evolution of
most  main  sequence (MS)  stars  --  see \citet{althausetal10-1}  and
references therein.  Because nuclear reactions do not operate in their
deep interiors, the evolution of WDs can be considered as a relatively
simple gravothermal  cooling process.  The evolutionary  cooling times
are  now   very  accurate   \citep{fontaineetal01-1,  renedoetal10-1},
providing a reliable way of measuring  the WD age from the temperature
and surface gravity measured  observationally.  By taking into account
the   initial-to-final   mass  relationship   \citep{ferrarioetal05-1,
catalanetal08-1},  the  MS mass  of  the  WD  progenitor can  also  be
computed, and from it  the time spent on the MS  can be derived, using
the appropriate  evolutionary sequences.  Thus, one  can easily obtain
the total age  as the sum of  the MS lifetime of the  WD precursor and
the (cooling) age of the resulting WD that we observe today

A  large   fraction  of   MS  stars  are   found  in   binary  systems
\citep{raghavanetal10-1,  yuanetal15-1}.  Like  in  single stars,  the
vast majority  of primary  (or more massive)  stars in  these binaries
will also  become WDs.  There are  two main channels that  lead to the
formation  of   these  WDMS  binaries   \citep{willems+kolb04-1}.   In
$\sim$75 per cent of the cases  the initial orbital separation is wide
enough  to allow  the evolution  of  the primary  star as  if it  were
single.   Hence, the  separations of  these WDMS  binaries are  either
roughly the same as the initial ones, or wider due to mass loss of the
WD progenitors that  result in the adiabatic expansion  of the orbits.
In the  remaining $\sim$25 per cent  of the cases the  initial orbital
separation  of the  MS  binary  is short  enough  so  that the  binary
undergoes  a   common  envelope   phase  \citep{webbink08-1}.    As  a
consequence,  the  orbital period  distribution  of  WDMS binaries  is
bi-modal, with close WDMS binaries that evolved though common envelope
being concentrated  at short orbital  periods, and wide  WDMS binaries
that  evolved  avoiding mass  transfer  interactions  at long  orbital
periods \citep{farihietal10-1, nebotetal11-1}.

Wide WDMS binaries  offer a promising methodology to  provide input to
determine the AMR.  Metallicities can  be directly determined from the
MS components in the  same way as it is done for  single MS stars, and
accurate ages  can be determined  from the observed properties  of the
WDs. Since the  two components in each binary are  coeval, the WDs and
MS stars have the same age, and consequently employing the accurate WD
ages provides  also that  of the  MS companions.   We have  started an
observational campaign  to measure MS  star metallicities and  WD ages
from a statistically large sample of  WDMS binaries.  In this paper we
present  the  techniques applied  to  derive  ages using  the  stellar
parameters  measured directly  from  the WD  spectra, the  metallicity
obtained from  the MS companion  spectra and, finally, we  discuss the
AMR we obtained using a pilot data-set of 23 WDMS binaries.

\begin{table*}
\caption{\label{t-param}  Names and  coordinates of  the 23  wide WDMS
  binaries  studied  in  this  work.  WD  stellar  parameters  (namely
  effective temperature, T$_\mathrm{eff}$, surface gravity, $\log$\,g,
  and  mass) and  M dwarf  spectral  types (Sp)  obtained fitting  the
  X-Shooter spectra are also provided.  WD ages (obtained assuming the
  initial-to-final mass relation  of \citealt{catalanetal08-1}) and MS
  star  [Fe/H] values  are  given in  the last  two  columns.  The  WD
  parameters   and   age   of    SDSSJ\,0138-0016   are   taken   from
  \citet{parsonsetal12-3}.  The WD  parameters of SDSSJ\,0325-0111 are
  derived      fitting     the      available     SDSS      spectrum.}
\setlength{\tabcolsep}{2ex} \centering
\begin{small}
\begin{tabular}{ccccccccc}
\hline
\hline
Object & ra  & dec & T$_\mathrm{eff}$ (WD) & $\log$\,g (WD) & Mass (WD) & Sp (MS)& Age & [Fe/H]  \\
       & (degrees) & (degrees) & (K)  & (dex)  &  \Msun&   & Gyr  & dex \\
\hline
 SDSSJ\,0003-0503 &  0.98723   &  -5.05909    & 19967$\pm$131  &  8.07$\pm$0.02 &  0.66$\pm$0.01 & M4  & 0.75$^{+0.13}_{-0.15}$   &   0.05 $\pm$  0.12 \\
 SDSSJ\,0005-0544 &  1.49948   &  -5.73780    & 32748$\pm$224  &  7.73$\pm$0.03 &  0.54$\pm$0.01 & M2  & 6.71$^{+2.22}_{-2.22}$   &   0.06 $\pm$  0.12 \\
 SDSSJ\,0036+0700 &  9.01079   &   7.01311    & 36105$\pm$ 49  &  7.87$\pm$0.02 &  0.60$\pm$0.01 & M4  & 1.52$^{+0.22}_{-0.30}$   &   0.30 $\pm$  0.12 \\
 SDSSJ\,0052-0051 & 13.03508   &  -0.85961    & 11933$\pm$ 94  &  8.02$\pm$0.03 &  0.61$\pm$0.02 & M4  & 1.34$^{+0.17}_{-0.31}$   &   0.06 $\pm$  0.12 \\
 SDSSJ\,0111+0009 & 17.84954   &   0.15981    & 12326$\pm$ 47  &  7.76$\pm$0.03 &  0.54$\pm$0.02 & M3  & 5.06$^{+1.74}_{-1.74}$   &  -0.46 $\pm$  0.12 \\
 SDSSJ\,0138-0016 & 24.71475   &  -0.27267    &  3570$\pm$110  &  7.92$\pm$0.02 &  0.54$\pm$0.01 & M5  & 9.50$^{+0.30}_{-0.20}$   &  -0.56 $\pm$  0.12 \\
 SDSSJ\,0256-0730 & 44.04421   &  -7.50683    & 10194$\pm$ 68  &  8.84$\pm$0.04 &  1.12$\pm$0.01 & M5  & 2.96$^{+0.20}_{-0.20}$   &  -0.27 $\pm$  0.12 \\
 SDSSJ\,0258+0109 & 44.57446   &   1.16278    & 36873$\pm$123  &  7.75$\pm$0.02 &  0.55$\pm$0.01 & M3  & 4.87$^{+1.16}_{-1.16}$   &   0.29 $\pm$  0.12 \\
 SDSSJ\,0321-0016 & 50.40225   &  -0.27511    & 31096$\pm$ 32  &  7.88$\pm$0.02 &  0.59$\pm$0.01 & M5  & 1.73$^{+0.26}_{-0.34}$   &   0.17 $\pm$  0.12 \\
 SDSSJ\,0325-0111 & 51.29517   &  -1.18725    & 10499$\pm$ 14  &  8.13$\pm$0.04 &  0.68$\pm$0.02 & M2  & 1.10$^{+0.12}_{-0.15}$   &  -0.36 $\pm$  0.12 \\
 SDSSJ\,0331-0054 & 52.88383   &  -0.91483    & 30742$\pm$ 30  &  7.96$\pm$0.01 &  0.63$\pm$0.01 & M3  & 0.97$^{+0.05}_{-0.09}$   &   0.09 $\pm$  0.12 \\
 SDSSJ\,0824+1723 & 126.1209   &   17.3959    & 12476$\pm$ 52  &  7.86$\pm$0.03 &  0.54$\pm$0.02 & M3  & 5.22$^{+1.94}_{-1.94}$   &  -0.10 $\pm$  0.12 \\
 SDSSJ\,0832-0430 & 128.2300   &  -4.51285    & 16064$\pm$ 85  &  8.01$\pm$0.01 &  0.62$\pm$0.01 & M1  & 0.94$^{+0.05}_{-0.06}$   &  -0.76 $\pm$  0.12 \\
 SDSSJ\,0916-0031 & 139.0061   &  -0.52494    & 19130$\pm$ 63  &  8.30$\pm$0.02 &  0.81$\pm$0.01 & M4  & 0.44$^{+0.02}_{-0.02}$   &   0.30 $\pm$  0.12 \\
 SDSSJ\,0933+0926 & 143.2996   &   9.44508    & 30401$\pm$ 18  &  7.63$\pm$0.01 &  0.54$\pm$0.01 & M5  & 6.05$^{+0.89}_{-0.89}$   &  -0.08 $\pm$  0.12 \\
 SDSSJ\,1023+0427 & 155.8927   &   4.45617    & 20498$\pm$ 68  &  7.89$\pm$0.02 &  0.57$\pm$0.01 & M4  & 2.79$^{+0.50}_{-0.67}$   &   0.18 $\pm$  0.12 \\
 SDSSJ\,1040+0834 & 160.2395   &   8.57267    & 10254$\pm$  8  &  8.00$\pm$0.03 &  0.60$\pm$0.02 & M5  & 1.62$^{+0.22}_{-0.44}$   &  -0.09 $\pm$  0.12 \\
 SDSSJ\,1405+0409 & 211.3955   &   4.15183    & 20716$\pm$ 88  &  8.15$\pm$0.02 &  0.71$\pm$0.01 & M4  & 0.46$^{+0.10}_{-0.11}$   &  -0.30 $\pm$  0.12 \\
 SDSSJ\,1527+1007 & 231.9337   &   10.1229    & 34079$\pm$ 52  &  7.86$\pm$0.02 &  0.59$\pm$0.01 & M3  & 1.42$^{+0.23}_{-0.30}$   &  -0.17 $\pm$  0.12 \\
 SDSSJ\,1539+0922 & 234.8943   &   9.37265    & 11183$\pm$143  &  8.72$\pm$0.03 &  1.06$\pm$0.01 & M5  & 2.49$^{+0.20}_{-0.20}$   &   0.29 $\pm$  0.12 \\
 SDSSJ\,1558+0231 & 239.7217   &   2.52731    & 30062$\pm$  9  &  7.79$\pm$0.01 &  0.54$\pm$0.01 & M4  & 5.57$^{+0.63}_{-0.63}$   &   0.07 $\pm$  0.12 \\
 SDSSJ\,1624-0022 & 246.1319   &  -0.38006    & 26291$\pm$203  &  7.92$\pm$0.03 &  0.60$\pm$0.01 & M3  & 1.26$^{+0.24}_{-0.36}$   &   0.02 $\pm$  0.12 \\
 SDSSJ\,2341-0947 & 355.4926   &  -9.78794    &  9433$\pm$ 26  &  8.27$\pm$0.04 &  0.77$\pm$0.03 & M4  & 1.45$^{+0.09}_{-0.10}$   &   0.07 $\pm$  0.12 \\
 \hline
\end{tabular} 
\end{small}
\end{table*}

\section{The WDMS binary sample}

Our  sample is  the largest  and  most homogeneous  catalogue of  WDMS
binaries  currently known  \citep{rebassa-mansergasetal16-1}, obtained
from the  Sloan Digital  Sky Survey  (SDSS).  Radial  velocity studies
have allowed  the identification of  hundreds of both close  SDSS WDMS
binaries  that  evolved  through  a common  envelope  phase  and  wide
binaries  that  did   not  interact  \citep{rebassa-mansergasetal07-1,
rebassa-mansergasetal11-1,  schreiberetal08-1}.   Close WDMS  binaries
cannot  be considered  in  our program,  as the  evolution  of the  WD
progenitors was truncated  during the common envelope  phase and hence
these WDs do not follow a clear initial-to-final mass relation.  Among
the sub-population of wide SDSS WDMS binaries, we selected 118 systems
which are  bright enough ($g<19$\,mag)  to allow deriving  accurate WD
ages  and MS  star  metallicities from  spectroscopic observations  at
current medium- and large-aperture ground based telescopes.

\section{Observations}
\label{s-obs}

We observed  19 wide WDMS binaries  in our sample with  the Very Large
Telescope  at  Cerro  Paranal  (Chile)  equipped  with  the  X-Shooter
instrument \citep{vernetetal11-1} -- the  target names and coordinates
are  provided in  Table\,\ref{t-param}.  X-Shooter  takes simultaneous
spectra   on   three   different  arms   (UVB;   3000--5600\AA,   VIS;
5500--10200\AA,    NIR;    10200--24800\AA)    thus    covering    the
3000--24\,800\AA\, wavelength  range in one single  exposure.  We used
the 0.9-1''  slits, which resulted  in resolving powers of  4350, 7450
and 5300 in the UVB, VIS and NIR arms, respectively.  The observations
were performed in service mode  throughout 2015 and the exposure times
were chosen  so that the  resulting spectra were of  a signal-to-noise
(SN) ratio of $\simeq$40 in the  regions of interest (the Balmer lines
in the UVB/VIS and the  \Ion{Na}{I} absorption doublet at 2205/2209 nm
in   the  NIR).    A  log   of   the  observations   is  provided   in
Table\,\ref{t-log}.  In  the NIR  the spectra  were observed  in stare
mode.   The   data  were  reduced,  wavelength   calibrated  and  flux
calibrated with the $esoreflex$ X-Shooter pipeline, version 2.6.8.  An
example of reduced and calibrated spectrum  is shown in the left panel
of Figure\,\ref{f-spectrum}.

Three additional wide SDSS WDMS binaries in our sample were previously
observed by  X-Shooter by some of  us in a separate  program, aimed at
measuring  accurate temperatures  for  ZZ\,Ceti WDs  in detached  WDMS
binaries  \citep{pyrzasetal15-1}:  SDSSJ\,0052-0051,  SDSSJ\,0824+1723
and SDSSJ\,0111+0009 (Table\,\ref{t-param}).  We reduced these data in
the same way as described above.

Finally,  we included  in our  sample the  eclipsing SDSS  WDMS binary
SDSSJ\,0138-0016  \citep{parsonsetal12-3} --  see Table\,\ref{t-param}
-- which  has  also  been  observed  intensively  with  the  X-Shooter
instrument.   With an  orbital period  of 1.7  hours, this  system has
certainly  evolved  through  a  common envelope  phase.   However,  it
contains  an ultra-cool  WD (3570  K)  that allows  deriving a  robust
cooling age  of 9.5 Gyr. We  reduced and calibrated the  NIR X-Shooter
data  of  SDSSJ\,0138-0016  to  derive   the  metallicity  of  the  MS
companion.

The  X-Shooter  spectra of  all  our  targets except  SDSSJ\,0325-0111
clearly  displayed  the WD  and  MS  features necessary  for  deriving
accurate ages  and metallicities (see  a description of our  method in
Section\,\ref{s-results}).   SDSSJ\,0325-0111  turned   out  to  be  a
spatially resolved WDMS binary in  the X-Shooter acquisition image for
which the  X-Shooter spectrum  only displayed  the MS  component, i.e.
the  WD ended  up outside  the  slit.  We  identified five  additional
spatially resolved binaries among our  23 targets as revealed by their
X-Shooter  acquisition   images:  SDSSJ\,0832-0430,  SDSSJ\,0933+0926,
SDSSJ\,1405+0409,  SDSSJ\,1558+0231 and  SDSSJ\,2341-0947. This  could
affect  the  relative  flux  contribution  of the  two  stars  in  the
X-Shooter spectra,  as the  probability exists that  they did  not fit
both   fully  into   the  slit.    However,   as  we   will  show   in
Section\,\ref{s-results}, we  do not  make use  of absolute  fluxes to
derive the ages and metallicities, hence this issue is not expected to
affect our results.

\begin{table}
\setlength{\tabcolsep}{0.6ex}
\caption{\label{t-log} Log of the observations. The dates are provided
  in the  second column. The exposure  times in each arm  are given in
  the last  three columns in  the format $n  \times t$, where  $n$ is
  number of  spectra.  We  average the  spectra before  performing the
  analysis.   The   spectra  of   SDSS\,J0052-0051,  SDSS\,J0111+0009,
  SDSS\,J0138-0016  and  SDSS\,J0824+1723  were obtained  as  part  of
  different programs but were  independently reduced and calibrated by
  us in the same way as our own data (Section\,\ref{s-obs}).}
\begin{flushleft}
\begin{center}
\begin{tabular}{ccccc}\hline\hline
  Object & Date/s & Exp. time  & Exp. time & Exp. time  \\
         &        &  (UVB)     &  (VIS)    &  (NIR)  \\
         &        &  n $\times$ seconds   & n $\times$ seconds   & n $\times$ seconds \\
\hline
SDSS\,J0003-0503	& 2015-09-25 &	1$\times$1240 & 1$\times$1200 &  5$\times$245 \\	
SDSS\,J0005-0544	& 2015-10-01 &	1$\times$1240 & 1$\times$1190 &  5$\times$245 \\ 	
SDSS\,J0036+0700	& 2015-10-01 &	 1$\times$430 &  1$\times$380 &  2$\times$220 \\ 	
SDSS\,J0052-0051  & 2011-10-27 &	4$\times$1475 & 4$\times$1420 &  8$\times$600 \\	
SDSS\,J0111+0009  & 2011-10-24 &	4$\times$1475 & 4$\times$1420 &  8$\times$600 \\	
           	  & 2011-10-28 &         &        &        \\
SDSS\,J0138-0016  & 2011-12-25 &	12$\times$606 & 24$\times$294 & 80$\times$100 \\	
SDSS\,J0256-0730	& 2015-09-25 &	 1$\times$450 &  1$\times$400 &  2$\times$230 \\ 	
SDSS\,J0258+0109	& 2015-09-18 &	1$\times$1240 & 1$\times$1190 &  5$\times$245 \\	
SDSS\,J0321-0016	& 2015-09-25 &	 1$\times$860 &  1$\times$810 &  3$\times$285 \\ 	
SDSS\,J0325-0111	& 2015-09-18 &	 1$\times$430 &  1$\times$375 &  2$\times$220 \\ 	
SDSS\,J0331-0054	& 2015-08-25 &	1$\times$1080 & 1$\times$1065 &  4$\times$275 \\ 	
SDSS\,J0824+1723  & 2011-12-23 &  4$\times$1475 & 4$\times$1420 & 14$\times$600 \\	
SDSS\,J0832-0430	& 2015-04-03 &	 2$\times$660 &  2$\times$610 &  5$\times$220 \\ 	
SDSS\,J0916-0031	& 2015-04-03 &	 1$\times$850 &  1$\times$830 &  3$\times$280 \\ 	
SDSS\,J0933+0926	& 2015-04-30 &	 1$\times$860 &  1$\times$805 &  3$\times$285 \\ 	
SDSS\,J1023+0427	& 2015-05-03 &	 1$\times$860 &  1$\times$805 &  3$\times$285 \\ 	
SDSS\,J1040+0834	& 2015-05-03 &	1$\times$1240 & 1$\times$1190 &  5$\times$245 \\ 	
SDSS\,J1405+0409	& 2015-04-29 &	1$\times$1640 & 1$\times$1595 &  6$\times$270 \\ 	
SDSS\,J1527+1007	& 2015-04-05 &	 1$\times$205 &  1$\times$150 &  1$\times$220 \\ 	
SDSS\,J1539+0922	& 2015-04-05 &	1$\times$1500 & 1$\times$1480 &  6$\times$250 \\ 	
SDSS\,J1558+0231	& 2015-04-29 &	 1$\times$280 &  1$\times$270 &  1$\times$285 \\ 	
SDSS\,J1624-0022	& 2015-04-05 &	1$\times$1500 & 1$\times$1480 &  6$\times$250 \\	
SDSS\,J2341-0947	& 2015-06-04 &	 1$\times$860 &  1$\times$805 &  6$\times$285 \\ 	
		& 2015-10-01 &         &        &        \\
\hline
\end{tabular}
\end{center}
\end{flushleft}
\end{table}

\section{Results}
\label{s-results}

In total we have X-Shooter spectra  of 23 wide WDMS binaries.  Here we
report the  spectral analysis performed to  derive the WD ages  and MS
star  metallicities together  with  the AMR  derived  from this  pilot
sample.

\subsection{WD ages}

We    used    the    decomposition/fitting   routine    outlined    by
\citet{rebassa-mansergasetal07-1} to subtract the MS star contribution
from the combined UVB and VIS  arm X-Shooter spectra and record the MS
star spectral type.  We then fitted the normalised Balmer lines of the
residual WD spectra (see an example in Figure\,\ref{f-spectrum}, right
panel) with the  model grid of \citet{koester10-1}  considering the 3D
corrections  by  \cite{tremblayetal13-1},  and  derived  WD  effective
temperatures and  surface gravities.  We interpolated  these values on
the cooling tracks  of \citet{renedoetal10-1} to obtain  the WD masses
and cooling ages.  We then  used the initial-to-final mass relation of
\cite{catalanetal08-1} to  derive the WD progenitor  masses.  Finally,
the  WD  progenitor  lifetimes  were  obtained  interpolating  the  WD
progenitor       masses       in      the       BASTI       isochrones
\citep{pietrinfernietal04-1}, for  which we adopted  the metallicities
as derived  from the MS  companions, see  below.  The WD  cooling ages
added to the MS lifetimes of  their progenitors gave the total ages of
the binaries. The  age uncertainties were obtained  propagating the WD
mass  uncertainties  resulting  from  the fits  (typical  errors  were
$\sim0.01$\Msun,  see Table\,\ref{t-param}).   The  WD  ages are  also
reported in Table\,\ref{t-param}.   Inspection of Table\,\ref{t-param}
reveals  that   the  age   uncertainties  increase   considerably  for
decreasing  WD  masses.   This  can   be  explained  as  follows:  the
progenitors of  low-mass WDs of  mass $<0.57\Msun$ are  low-mass stars
which spend long  time in the MS.   This time spent in the  MS is very
sensitive to the  mass of the star,  thus even small errors  in the WD
progenitor masses  translate into rather  different ages spent  on the
MS.

As  mentioned  in  Section\,\ref{s-obs},  the  X-Shooter  spectrum  of
SDSSJ\,0325-0111  displayed only  the  features of  the MS  component,
hence  no  WD parameters  could  be  measured.   Thus, we  fitted  the
available optical  SDSS spectrum  of this  object in  the same  way as
described  above  to  measure  the   WD  stellar  parameters  and  the
age\footnote{All  SDSS  WDMS  binaries   studied  in  this  work  have
available SDSS  spectra.  However, the  SN ratio of these  spectra are
generally  low  ($\la$10-15; see  \citealt{rebassa-mansergasetal16-1})
and  the derived  WD  parameters are  subject  to considerably  larger
uncertainties than those  measured from the X-Shooter  spectra of much
higher SN  ratio.  Fortunately, the SDSS  spectrum of SDSSJ\,0325-0111
was  an exception  (with a  SN$=$40) and  we could  derive precise  WD
parameters, hence age.}.

\begin{figure}
\centering
\includegraphics[angle=-90,width=\columnwidth]{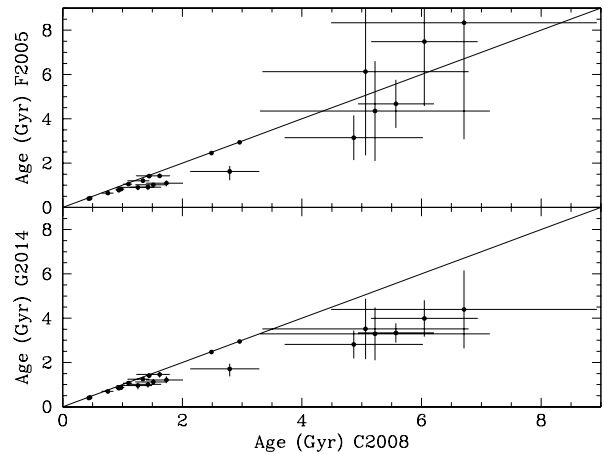}
\caption{Comparison between the WD  ages derived using three different
  initial-to-final  mass  relations:  the semi-empirical  relation  of
  \citet{catalanetal08-1} adopted in this work labelled as C2008 (this
  relation  is  virtually identical  to  the  theoretical relation  of
  \citealt{renedoetal10-1}),     the     empirical     relation     of
  \citet{ferrarioetal05-1}  (F2005)  and  the  empirical  relation  of
  \citet{gesickietal14-1} (G2014).}
\label{f-ages}
\end{figure}

This method of calculating total ages as the sum of the WD cooling and
the MS  progenitor lifetimes  has been  tested using  WDs in  open and
globular clusters, with the WD ages derived in this way and those from
the    MS   turn-off    of   the    clusters   being    very   similar
\citep{garcia-berroetal13-1,   torresetal15-1}.     However,   it   is
important  to   mention  that  these  works   employ  the  theoretical
initial-to-final  mass relation  of \citet{renedoetal10-1}.   Although
this theoretical relation is virtually identical to the semi-empirical
relation  from \citet{catalanetal08-1}  assumed  in this  work, it  is
important to bear in mind  that the initial-to-final mass relation for
WDs  remains still  rather  unconstrained  observationally.  We  hence
decided to  re-derive our  ages using  two additional  (and empirical)
initial-to-final    relations,     namely    those     presented    by
\citet{ferrarioetal05-1} and  \citet{gesickietal14-1}. As can  be seen
in Figure\,\ref{f-ages},  we found no substantial  differences between
the   ages  obtained   using  these   relations  and   the  one   from
\citet{catalanetal08-1}.

\subsection{MS star metallicities}

Due  to selection  effects,  the  SDSS WDMS  binary  sample is  biased
towards the  detection of  low-mass MS companions  of spectral  type M
\citep{rebassa-mansergasetal10-1}.    Several    methods   exist   for
measuring  M  dwarf  metallicities (specifically  [Fe/H]  abundances),
either from high-resolution  optical \citep{nevesetal14-1} or infrared
\citep{lindgrenetal16-1}  spectra,   or  from   low/medium  resolution
infrared   spectra    such   as   those   obtained    in   this   work
\citep{rojas-ayalaetal12-1,   mannetal14-1,    newtonetal14-1}.    The
methods      described     by      \citet{rojas-ayalaetal12-1}     and
\cite{mannetal14-1} make use of  semi-empirical relations based on the
equivalent  widths   of  different   atomic  lines   (\Ion{Ca}{I}  and
\Ion{Na}{I})  as  well  as  the  H$_2$0-K2  index  to  obtain  [Fe/H].
However, deriving  the H$_2$0-K2  index requires measuring  the median
flux in the  2360--2480 nm range, which is  unfortunately dominated by
noise in  our spectra.  Thus,  we obtained the [Fe/H]  abundances from
the   K-band,   NIR  X-Shooter   spectra   of   the  M   dwarfs   (see
Figure\,\ref{f-spectrum})   following  the   procedure  described   in
\citet{newtonetal14-1}.   This  method  provides  [Fe/H]  following  a
semi-empirical  multivariate linear  regression  based  solely on  the
\Ion{Na}{I}  absorption   doublet  (2205/2209\,nm)   equivalent  width
(EWNa).  The \Ion{Na}{I}  doublet has been proven to be  a good tracer
of metallicity  \citep[e.g.][]{coveyetal10-1}.  \citet{newtonetal14-1}
calibrated their metallicity relation using  M dwarfs in common proper
motion  pairs with  F, G,  K stars  and demonstrated  that the  [Fe/H]
values obtained  in this way are  accurate up to 0.12  dex.  They also
claimed that  the relation is  well behaved  for M dwarfs  of spectral
types M1--M5, which is the case for  all M dwarfs in our observed WDMS
binaries  (Table\,\ref{t-param}).  The  validity  of  the relation  of
\citet{newtonetal14-1}  has  been  tested  by  \citet{veyetteetal16-1}
using  PHOENIX   atmospheric  models,   who  concluded  that   it  was
appropriate.

In order to measure the EWNa values we corrected the systemic (radial)
velocities of the binaries and normalised the flux of each spectrum in
the 2194--2220\,nm  region fitting a third-order  spline function. The
\Ion{Na}{I}   doublet  feature   was  excluded   in  the   process  of
normalisation.  We  then used  the trapezoidal  rule to  integrate the
flux  of absorption  doublet  within the  2204--2210  nm region.   The
values of [Fe/H] we derived are listed in Table\,\ref{t-param}.

In the  upper panel  of Figure\,\ref{f-amr}  we display  our resulting
AMR. In  the bottom panel  of the same  figure we display  the average
$\langle$[Fe/H]$\rangle$  per 1~Gyr  bins  along  with their  standard
deviations   ($\sigma$).    These   values  are   also   reported   in
Table\,\ref{t-amr}.  The  AMR derived in  this way shows  an intrinsic
scatter  $>$0.2\,dex  for  most  ages, independently  of  the  assumed
initial-to-final mass  relation.  This significant scatter  suggests a
lack of correlation between the values of [Fe/H] and ages derived from
our pilot data-set of 23 WDMS systems.

\section{Discussion}

The existence of an AMR in the solar vicinity has been long discussed.
Earlier studies of field stars  displayed a trend of decreasing [Fe/H]
for   increasing   ages  \citep{twarog80-1,   edvardsson93-1}.    More
recently,   \citet{rocha-pintoetal00-1}  and   \cite{soubiranetal08-1}
found also  a correlation between  age and [Fe/H]  using chromospheric
activity to derive the ages of  late-type stars and ages derived using
isochrones  for  a sample  of  giants,  respectively. However,  recent
results  using  mainly  a  sample  of  turn-off  stars  together  with
isochrone fitting techniques to derive stellar ages show a substantial
scatter in the relation as well as a nearly flat distribution for ages
younger  than  $\sim$6-8\,Gyr,  suggesting  there  is  no  correlation
between    age    and    metallicity    in    the    solar    vicinity
\citep{feltzingetal01-1,     holmbergetal09-1,     casagrandeetal11-1,
  haywoodetal13-1, bergemannetal14-1}.   Using high-resolution spectra
of solar  twins, \citet{nissen15-1} found  also a lack  of correlation
between [Fe/H]  and age over an  age interval of 8~Gyr.   However, for
many  of  the  elements  \citet{nissen15-1} found  there  is  a  tight
correlation  between [X/Fe]  and  stellar age  with  amplitudes up  to
$\sim$0.15\,dex.

\begin{figure}
\centering
\includegraphics[angle=-90,width=\columnwidth]{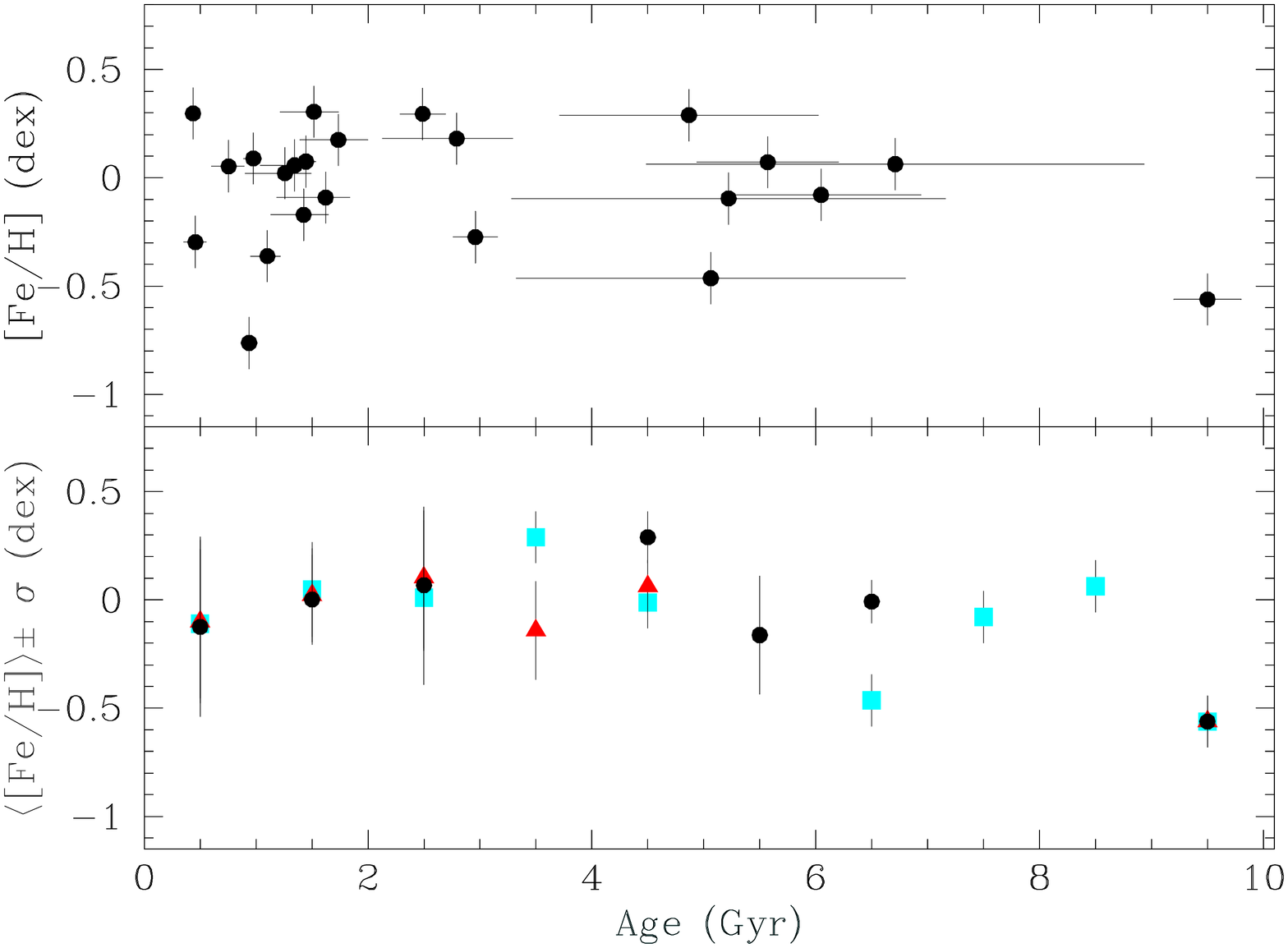}
\caption{Upper panel: the age-[Fe/H] relation derived from the 23 SDSS
  WDMS binaries studied in this  work.  The ages are obtained assuming
  the  initial-to-final  mass   relation  of  \citet{catalanetal08-1}.
  Bottom panel:  average [Fe/H] values  for 1 Gyr bins,  together with
  their  standard deviations  ($\sigma$)  for ages  derived using  the
  initial-to-final mass  relation of \citet{catalanetal08-1}  -- black
  dots   --   \citet{ferrarioetal05-1}   --  cyan   squares   --   and
  \citet{gesickietal14-1} -- red triangles.  In those cases where only
  one point  falls within the age  bin, we assume $\sigma$  to be 0.12
  dex,   i.e.    the  [Fe/H]   accuracy   of   the  method   we   used
  \citep{newtonetal14-1}.  The  average [Fe/H] values are  provided in
  Table\,\ref{t-amr}.   Independently  of  the  initial-to-final  mass
  relation used, the average values of [Fe/H] are nearly identical for
  the  0.5,  1.5 and  2.5  Gyr  age  bins, and  become  systematically
  different for  larger ages  (see Table\,\ref{t-amr}).   The observed
  scatter of [Fe/H] at all  ages remains, independently of the adopted
  initial-to-final mass relationship.}
\label{f-amr}
\end{figure}

The discrepancies observed  in the AMR reported above  are most likely
due  to the  difficulty  of  measuring precise  ages  for field  stars
\citep{soderblom10-1}.  In order to overcome  this problem the AMR has
been analysed  using open clusters  \citep{friel95-1, carraroetal98-1,
pancinoetal10-1}.  These studies support a lack of correlation between
age  and [Fe/H].   However,  it  is important  to  emphasise that  the
lifetimes   of    open   clusters    are   generally    below   1\,Gyr
\citep{vandeputteetal10-1}, which makes it  difficult to constrain the
AMR at  intermediate and  old ages.  Moreover,  the current  sample of
open clusters has not been analysed in an homogeneous way, which would
leave these studies inconclusive \citep{casamiquelaetal16-1}.

In order  to provide  additional observational input  we used  in this
work  23  wide binaries  (except  SDSSJ0138-0016,  an eclipsing  close
binary  of known  age) composed  of a  WD and  a MS  star to  test the
existence of  a correlation between  age and metallicity in  the solar
vicinity.  Our  analysis clearly  illustrates the lack  of correlation
between age and  [Fe/H] at young and intermediate ages  (0--7 Gyr, see
Figure\,\ref{f-amr}).    \citet{zhaoetal11-1}   performed  a   similar
exercise using  a sample  of 21 wide  WDMS binaries  mostly containing
F-,~G- or K-type companions.  Their results suggested the existence of
a correlation between age and metallicity.  However, to derive WD ages
they used  cooling models \citep[e.g.][]{wood95-1}, as  well as pre-WD
evolutionary   times    \citep{iben+laughlin89-1}   that    were   not
state-of-the-art.

For  young  and intermediate  ages  ($\la$8\,Gyr),  we find  that  the
averaged AMR fluctuates between $-0.4$ to +0.3\,dex, but does not show
any apparent slope.  Although more data  are needed to draw more solid
conclusions  about  the   slope  of  the  AMR  in   this  region,  our
observations  agree  with similar  studies  \citep{casagrandeetal11-1,
  haywoodetal13-1,  bergemannetal14-1}  using sub-giants  and  turnoff
stars.  Also, \citet{casagrandeetal11-1}, \citet{haywoodetal13-1}, and
\citet{bergemannetal14-1} found  a decline  of [Fe/H] for  ages longer
than  $\sim$8\,Gyr. However,  our  current  data do  not  allow us  to
confirm, nor  discard, this  trend of  the AMR.   Finally, we  find an
intrinsic scatter  of $\sim$0.2\,dex in  our AMR, a value  larger than
the nominal observational error on  [Fe/H] (0.12\,dex).  The fact that
we observe a large scatter of  [Fe/H] for young and intermediate ages,
as  previously   reported  in  several  other   observational  studies
targeting  field stars,  supports the  idea  of the  methods used  for
calculating  stellar  ages  (e.g.  chromospheric  activity,  isochrone
fitting) being  reliable.  However, it is  noteworthy that differences
arise when comparing  individual ages of selected  field stars derived
using  different  techniques   \citep{anguianoetal10-1}.   This  makes
difficult to assess which of the employed methods is more reliable.

\begin{table}
\setlength{\tabcolsep}{0.4ex}
\caption{\label{t-amr} $\langle$[Fe/H]$\rangle  \pm \sigma$  for 1~Gyr
  bins. In those cases where only  one point falls within the age bin,
  we assume $\sigma$ to be 0.12  dex, i.e.  the [Fe/H] accuracy of the
  method we  used \citep{newtonetal14-1}. The first  row indicates the
  initial-to-final mass relation used  for calculating the ages: C2008
  \citep{catalanetal08-1},  F2005  \citep{ferrarioetal05-1} and  G2014
  \citep{gesickietal14-1}.}
\begin{flushleft}
\begin{center}
\begin{tabular}{ccccccccc}\hline\hline
      C2008 &  &  &  F2005 & & & G2014 & & \\
Age bin & $\langle$[Fe/H]$\rangle$ & $\sigma$ & Age bin & $\langle$[Fe/H]$\rangle$ & $\sigma$ & Age bin & $\langle$[Fe/H]$\rangle$ & $\sigma$ \\
  (Gyr) &   (dex)      &  (dex)   \\
\hline
       0.5    &  $-0.12$  &     0.42  &   0.5  &    $-0.11$  &   0.34  &     0.5  &    $-0.10$ &      0.38\\
       1.5    &     0.00  &     0.21  &   1.5  &       0.05  &   0.22  &     1.5  &       0.02 &      0.22\\
       2.5    &   0.07  &       0.30  &   2.5  &       0.01  &   0.40  &     2.5  &       0.10 &      0.33\\
       3.5    &   ---    &      ---   &   3.5  &       0.29  &   0.12  &     3.5  &    $-0.14$ &      0.23\\
       4.5    &   0.29  &       0.12  &   4.5  &    $-0.01$  &   0.12  &     4.5  &       0.06 &      0.12\\
       5.5    &  $-0.16$  &     0.27  &   5.5  &     ---     &   ---   &     5.5  &     ---    &      --- \\
       6.5    &  $-0.01$  &     0.10  &   6.5  &    $-0.46$  &   0.12  &     6.5  &     ---    &      --- \\
       7.5    &   ---     &     ---   &   7.5  &    $-0.08$  &   0.12  &     7.5  &     ---    &      --- \\
       8.5    &   ---     &     ---   &   8.5  &       0.06  &   0.12  &     8.5  &     ---    &      --- \\
       9.5    &  $-0.56$  &     0.12  &   9.5  &    $-0.56$  &   0.12  &     9.5  &    $-0.56$ &      0.12\\
\hline
\end{tabular}
\end{center}
\end{flushleft}
\end{table}

Our results  provide clear  additional observational evidence  for the
existence of a  physical mechanism/s that causes  the observed scatter
of [Fe/H]  in the  observed AMR in  the solar  neighbourhood.  Several
mechanisms have been suggested to  explain the scatter observed in the
AMR.  Among  them we  mention self-enrichment of  gas in  star forming
regions  \citep{pilyugin+edmunds96-1}, episodic  gas in-fall  onto the
disc  \citep{koppen+hensler05-1},  or   the  currently  most  accepted
scenario, invoking  radial migration effects --  metal-rich stars form
in the inner  disc and subsequently migrate to  the metal-poorer outer
disc \citep{selwood+binney02-1, roskaretal08-1, minchevetal10-1}.  Due
to the small WDMS binary sample size we cannot confirm nor discard any
of these (or any other) scenarios.   Clearly, the analysis of a larger
sample will help in settling all these issues.

\section{Conclusions}

In  this work  we used  WDs  in WDMS  binaries  as tools  to test  the
existence of the AMR in the solar vicinity. The total ages for the WDs
were  obtained using  the  most  reliable cooling  tracks  as well  as
reliable   evolutionary   times   for   their   progenitors,   whereas
metallicities  for  MS stars  were  derived  using the  most  commonly
employed techniques.   This procedure  has allowed  us to  derive both
accurate ages and  reliable metallicities in an  homogeneous way.  Our
analysis of a  pilot sample of 23 of such  systems clearly illustrates
the  lack  of  correlation  between   age  and  [Fe/H]  at  young  and
intermediate   ages  (0--7   Gyr).    This   result  provides   robust
observational  evidence for  the existence  of a  physical mechanism/s
that causes the observed scatter of [Fe/H] in the observed AMR.

\section*{Acknowledgments}

This research has been funded  by MINECO grant AYA2014-59084-P, by the
AGAUR, by  the European  Research Council  under the  European Union's
Seventh   Framework  Programme   (FP/2007-2013)/ERC  Grant   Agreement
n.320964 (WDTracer), and by the National Key Basic Research Program of
China (2014CB845700). BA gratefully acknowledges the financial support
of the  Australian Research  Council through Super  Science Fellowship
FS110200035.

Based on observations made with ESO Telescopes at the La Silla Paranal
Observatory  under programme  ID\,095.B-0019. Based  on data  products
from observations  made with  ESO Telescopes at  the La  Silla Paranal
Observatory under programmes ID\,088.D-0481 and 288.D-5015.

\end{document}